# Spinons in a new Shastry-Sutherland lattice magnet Pr$_2$Ga$_2$BeO$_7$

N. Li[1,14], A. Brassington[2,14], M. F. Shu[3,14], Y. Y. Wang[1], H. Liang[1], Q. J. Li[4], X. Zhao[5], P. J. Baker[6], H. Kikuchi[7], T. Masuda[7,8], G. Duan[9], C. Liu[10], H. Wang[11], W. Xie[11], R. Zhong[12], J. Ma[3]★, R. Yu[9,13]★, H. D. Zhou[2]★, and X. F. Sun[1]★

[1]Anhui Key Laboratory of Magnetic Functional Materials and Devices, Institutes of Physical Science and Information Technology, Anhui University, Hefei, Anhui 230601, People's Republic of China

[2]Department of Physics and Astronomy, University of Tennessee, Knoxville, Tennessee 37996-1200, USA

[3]Key Laboratory of Artificial Structures and Quantum Control, Shenyang National Laboratory for Materials Science, School of Physics and Astronomy, Shanghai Jiao Tong University, Shanghai 200240, People's Republic of China

[4]School of Physics and Optoelectronics, Anhui University, Hefei, Anhui 230061, People's Republic of Chin

[5]School of Physics Sciences, University of Science and Technology of China, Hefei, Anhui 230026, People's Republic of China

[6]ISIS Facility, Rutherford Appleton Laboratory, Chilton, Didcot OX11 0QX, United Kingdom

[7]Institute for Solid State Physics, The University of Tokyo, Chiba 277-8581, Japan

[8]Institute of Materials Structure Science, High Energy Accelerator Research Organization, Ibaraki 305-0801, Japan

[9]Department of Physics and Beijing Key Laboratory of Opto-electronic Functional Materials and Micro-nano Devices, Renmin University of China, Beijing 100872, People's Republic of China

[10]School of Engineering, Dali University, Dali, Yunnan 671003, People's Republic of China

[11]Department of Chemistry, Michigan State University, East Lansing, Michigan 48824, USA

[12]Tsung-Dao Lee Institute, Shanghai Jiao Tong University, Shanghai, 201210, People's Republic of China

[13]Key Laboratory of Quantum State Construction and Manipulation (Ministry of Education), Renmin University of China, Beijing, 100872, People's Republic of China

[14]These authors contributed equally: N. Li, A. Brassington, M. F. Shu

*email: jma3@sjtu.edu.cn; rong.yu@ruc.edu.cn; hzhou10@utk.edu; xfsun@ahu.edu.cn

**Identifying the elusive spinon excitations in quantum spin liquid (QSL) materials is what scientists have long sought for. Recently, thermal conductivity ($\kappa$) has emerged to be a decisive probe because the fermionic nature of spinons leads to a characteristic nonzero linear $\kappa_0/T$ term while approaching zero Kelvin. So far, only a few systems have been reported to exhibit such term. Here, we report a $\kappa_0/T \approx 0.01$ WK$^{-2}$m$^{-1}$, the largest $\kappa_0/T$ value ever observed in magnetic oxide QSL candidates, in a new quantum magnet Pr$_2$Ga$_2$BeO$_7$ with a Shastry-Sutherland lattice (SSL). Its QSL nature is further supported by the power-law temperature dependence of the specific heat, a plateau of muon spin relaxation rate, and gapless inelastic neutron spectra. Our theoretical analysis reveals that the introduction of XY spin anisotropy is the key for Pr$_2$Ga$_2$BeO$_7$ to be the first QSL realized on the SSL, after more than four decades of extensive studies on this celebrated magnetically frustrated lattice.**

One of the most intriguing features of the extensively studied quantum spin liquid (QSL) state is the emergent fractionalized magnetic excitations, particularly the spinons[1-4]. Through decades of studies on QSL, the consensus of the community is that thermal conductivity is a decisive probe for the existence of spinons due to its only sensitivity to itinerant heat carriers. The exact reason is that the fermionic nature or the itinerant mobility of spinons can give rise to a nonzero $\kappa_0/T$ term in thermal conductivity in the zero temperature limit[5-9]. So far, only a few QSL candidates have been reported to exhibit such a nonzero $\kappa_0/T$ term, as listed in Table I[10-14]. Even though, the existence of spinons in some of these materials is still controversial when taking into account following issues: (i) the reported $\kappa_0/T$ value (nonzero or zero) is sample dependent; (ii) the reported value is pretty small due to spin-phonon scatterings; and (iii) the $\kappa_0/T$ term demises while approaching zero Kelvin as an effect of strong disorder, magnetic ordering, or possibly a gap opening.

Table I QSL candidates with nonzero $\kappa_0/T$ term reported.

| Material | Structure | $\kappa_0/T$ (WK$^{-2}$m$^{-1}$) | Note |
| --- | --- | --- | --- |
| EtMe$_3$Sb[Pd(dmit)$_2$]$_2$ | Triangular | 0.194 | sample dependent, zero $\kappa_0/T$ term reported[10,15,16] |
| 1$T$-TaS$_2$ | Triangular | 0.05 | sample dependent, zero $\kappa_0/T$ term reported[11,17] |
| PbCuTe$_2$O$_6$ | Hyperhyperkagome | 0.0075 | depletion of spinon transport below 0.3 K[12] |
| YbMgGaO$_4$ | Triangular | 0.0058 | spinon-phonon scattering, zero $\kappa_0/T$ term reported[13,18] |
| Na$_2$BaCo(PO$_4$)$_2$ | Triangular | 0.0062 | $T_N$ = 148 mK[14] |

Meanwhile, some other QSL candidates (as listed in Table SI of the Supplementary Information) exhibit supporting evidence of spinons, such as a power-law temperature dependence of the specific heat, a continuum in the spin excitation spectrum, and a plateau of the muon relaxation rate. But the absence of a $\kappa_0/T$ term questions the existence of spinons, and whether these materials are genuine QSLs or just QSL-like is therefore under debate.

As one of the most celebrated magnetically frustrated systems, the Shastry-Sutherland lattice (SSL) magnets have caught extensive attention for four decades, both theoretically and experimentally, and various exotic magnetic ground states have been observed[20-26]. However, so far, NO SSL magnet exhibits QSL behavior. In this paper, we first time studied a new quantum magnet $Pr_2Ga_2BeO_7$, in which the $Pr^{3+}$ ions form orthogonal dimers on the SSL[19]. Our thermal conductivity measurements on high quality single crystals reveal a $\kappa_0/T \approx 0.01$ WK$^{-2}$m$^{-1}$, the largest $\kappa_0/T$ value ever reported in magnetic oxide QSL candidates. Together with observations of the power-law temperature dependence of specific heat, a plateau of relaxation rate in muon spin relaxation ($\mu$SR) measurements, and the gapless spin excitations in inelastic neutron scattering (INS), our results reveal a gapless QSL behavior with dispersive spinon excitations in $Pr_2Ga_2BeO_7$. Furthermore, our theoretical analysis points out that the introduction of an XY-spin anisotropy is the key to stabilize such a QSL state which is realized experimentally for the first time in this new SSL magnet.

High quality single crystals of $Pr_2Ga_2BeO_7$ were first time grown using floating zone technique. The single crystal X-ray diffraction (SCXRD) data reveals that the magnetic $Pr^{3+}$ ions of $Pr_2Ga_2BeO_7$ locate on a SSL in the *ab* plane, which are separated by nonmagnetic Ga/Be-O polyhedra along the *c* axis, as shown in Fig. 1a. Here, the orthogonal dimers with intradimer coupling $J_1$ and interdimer coupling $J_2$ form the Shastry-Sutherland layer, as shown in Fig. 1b. The intradimer distance is 4.11817(0) Å, the interdimer distance is 3.60954(0) Å, and the interlayer one is 5.0693(14) Å. More XRD data and crystal structure parameters are shown in the Supplementary Information.

Figure 2 shows the low-temperature thermal conductivity ($\kappa$) of $Pr_2Ga_2BeO_7$ single crystal with heat current along the *a* or the *c* axis. Except for the broad phonon peak, $\kappa$ shows no sign of long-range magnetic order down to 70 mK. The most striking feature is that the linear fitting of the zero field $\kappa/T$ data gives the residual values of $\kappa_0/T = 0.0105$ and $0.0095$ WK$^{-2}$m$^{-1}$ for $\kappa_a$ and $\kappa_c$, respectively, as shown in Figs. 2b and 2d. The large residual $\kappa_0/T$ value prompts a significant fermionic contribution which most likely comes from spinons in this magnetic insulator. Here the data are fitted by $\kappa/T = a + bT$, where $a = \kappa_0/T$ representing the contribution from spinon excitations, and the second term is associated with both spinons and acoustic phonons. It is known that either spinons with a linear dispersion or spin-phonon scattering effect may give rise to a $T^2$ term in $\kappa$, which deviates from the conventional $T^3$ phononic contribution[13,14,27]. Two notable things as compared to other QSL materials: first, the observed $\kappa_0/T$ values are the largest among the studied magnetic oxide QSLs, as listed in Table I; second, the $T$ linear behavior of $\kappa$ persists down to the lowest temperature measured as 70 mK. For example, the reported $\kappa_0/T$ value of PbCuTe$_2$O$_6$[12] is comparable to those of Pr$_2$Ga$_2$BeO$_7$, however its linear behavior only exists above 0.3 K and below which the spinon transport quickly depletes with dropping thermal conductivity.

Similar data fittings for the 14 T data show that a high magnetic field smear out the residual thermal conductivity while enhancing $\kappa$. This is reasonable since the energy scale corresponding to this field may already exceed the bandwidth of spinons and hence completely suppress the fermionic spin excitations. In addition, the phonon term seems to be significantly enhanced in 14 T field, which can be understood as the weakened spin-phonon scattering. This indicates that in zero field, the spinon excitations not only transfer heat but also scatter phonons.

Below, we present several other observations that support the QSL nature of Pr$_2$Ga$_2$BeO$_7$. Figure 3a shows the temperature dependence of the inverse DC magnetic susceptibility $1/\chi(T)$ with an external field of 0.1 T along the $c$ or $a$ axis. No long-range magnetic order is observed down to 0.4 K for both field directions. The Curie-Weiss (CW) law fitting was performed at high temperature (50 − 300 K) and low temperature (0.8 − 8 K) ranges due to the influence of crystalline electric field (CEF) splitting, as indicated by red lines in Fig. 3a. The fitting to high temperature data yields the Curie-Weiss temperature $\theta_{CW}$ = − 9.79(8) K, the effective moment $\mu_{eff}$ = 3.18(3) $\mu_B$/Pr and $\theta_{CW}$ = − 35.83(7) K, $\mu_{eff}$ = 3.67(1) $\mu_B$/Pr for $B$ // $c$ and $B$ // $a$, respectively. A slope change of $1/\chi(T)$ was observed at low temperatures, and the low temperature CW fitting gives $\theta_{CW}$ = − 19.98(3) K, $\mu_{eff}$ = 3.77(3) $\mu_B$/Pr and $\theta_{CW}$ = − 21.34(1) K, $\mu_{eff}$ = 3.40(9) $\mu_B$/Pr for $B$ // $c$ and $B$ // $a$, respectively. The obtained effective moment is close to the theoretical free-ion value of Pr$^{3+}$ (3.58 $\mu_B$). The negative value of $\theta_{CW}$ obtained at low temperature fittings indicates the dominant antiferromagnetic (AFM) coupling between Pr$^{3+}$ spins and fits to the QSL scenario.

Figure 3b shows low-temperature specific heat data of Pr$_2$Ga$_2$BeO$_7$ single crystal at zero field, which exhibits no sign of a phase transition down to 50 mK. By comparing to the specific heat of La$_2$Ga$_2$BeO$_7$ (an isostructural and nonmagnetic material), it is notable that the low-temperature specific heat of Pr$_2$Ga$_2$BeO$_7$ is mainly the magnetic contribution and displays a $T^2$ dependence for 0.2 < $T$ < 3 K. Such a quadratic temperature dependence of magnetic specific heat has also been observed in several other QSL candidates[28-32] and was attributed to a gapless U(1) Dirac quantum spin liquid state, in which the spinons are gapless at certain Dirac nodes with a linearly dispersive spectrum[33,34]. Therefore, this feature again supports the existence of spinons in Pr$_2$Ga$_2$BeO$_7$. Moreover, the external magnetic field can significantly suppress the specific heat below 10 K, showing the predominately magnetic contribution at low temperatures (see Supplementary Information).

Figure 4 shows the $\mu$SR data measured on polycrystalline $Pr_2Ga_2BeO_7$. As shown in Fig. 4a, the absence of oscillations in the zero-field (ZF) signal and the lack of asymmetry recovery to 1/3 at high counting times rule out magnetic order or spin freezing in $Pr_2Ga_2BeO_7$ down to 30 mK. The time dependence of the muon decay asymmetry through the full temperature range (0.03 – 250 K) can be well described by Equation (1) listed in the Method. The fitted muon relaxation rates are plotted as a function of temperature in Fig. 4b. The $\lambda_{tail}(T)$, a weak exponential relaxation rate reflecting the depolarization due to the muon coupled to the nuclear magnetism, is nearly close to zero in the whole temperature range, corresponding to a simple exponential correction originating from the surrounding nuclei. Meanwhile, $\lambda_2$, the muon spin relaxation rate related to the dynamic internal magnetic fields, increases by more than 2 orders with decreasing temperature, showing a dynamical slowing down in $Pr_2Ga_2BeO_7$. Below 2 K, $\lambda_2$ reaches a plateau value, which persists down to 30 mK. Such a plateau in muon relaxation rate has been commonly observed in other QSL candidates[35-40].

As shown in Figure 5a, the elastic neutron scattering data in the (0KL) plane measured at 0.7 K exhibits no magnetic Bragg peaks, and again confirms the lack of long-range magnetic order in $Pr_2Ga_2BeO_7$. The INS spectrum measured at 0.7 K is presented in Fig. 5b and (c) as the spectral intensity along the high-symmetry momentum directions $R_1$-$Z_1$-$R_2$-$X_1$-$R_3$ and $Z_2$-$\Gamma_1$-$X_1$-$\Gamma_2$ in energy-momentum (E-Q) space. The main observation is that the spectral intensity is smeared in the whole Brillouin zone but with a gapless feature. Such gapless excitations support a gapless QSL state in $Pr_2Ga_2BeO_7$. Moreover, the largest intensity is found near the $Z_1$ point.

To understand the magnetism of $Pr_2Ga_2BeO_7$, we start from a theoretical analysis on the single-ion physics. $Pr^{3+}$ ion has a $4f^2$ electron configuration with a total angular momentum $J = 4$. Since the point group of the $Pr^{3+}$ site contains no rotational symmetry, the $J = 4$ non-Kramers multiplet split to 9 non-degenerate singlets under CEF, as verified by our point-charge model calculation (see Supplementary Information). Interestingly, as illustrated in Fig. 6a, we find the lowest two crystal field levels have a small splitting of about 0.215 meV and are well separated from other levels, which are located at least 60 meV above. Denoting these two states as $|\psi_+\rangle$ and $|\psi_-\rangle$, respectively, they constitute a quasi-degenerate non-Kramers doublet, and the magnetism of $Pr_2Ga_2BeO_7$ can be understood as an effective $S = 1/2$ XXZ-type Shastry-Sutherland model (SSM) (see Supplementary Information).

This model differs from the original SSM in several aspects: First, only $S^z$ transforms as a magnetic dipole, the $S^x$ and $S^y$ components transform as quadrupoles; Second, the model generically exhibits spin anisotropy $\Delta \neq 1$. For the Heisenberg SSM, strong spin frustration drives the plaquette-to-AFM transition close to a deconfined quantum critical point (DQCP)[41], but whether a QSL can be stabilized is still under debate[42]. In the XXZ SSM, we find, within density matrix renormalization group (DMRG) calculation, a state exhibiting neither plaquette nor AFM order but algebraically decaying spin correlation for $\Delta \geq 2.5$ (see Supplementary Information). This implies a gapless QSL ground state stabilized by increasing the spin anisotropy, as depicted in the phase diagram of Fig. 6b. This QSL-like state in our model provides a natural explanation to the experimental findings of absence of magnetic ordering in $Pr_2Ga_2BeO_7$.

As for the nature of this QSL, a likely scenario points to a Dirac QSL. Its linear spinon dispersion explains the $T^2$ behavior of the specific heat, and the unavoidable disorder in the system may cause a residual density of states in low energies, which gives rise to the finite $\kappa_0/T$ in the thermal conductivity (see Supplemental Information), in a way analogous to the case of $d$-wave superconductors[43].

In summary, the large residual $\kappa_0/T$ term in the thermal conductivity provides evidence for the existence of spinons in a new SSL magnet $Pr_2Ga_2BeO_7$. We further demonstrate that the introduction of XY-type spin anisotropy on SSL is the key for us to experimentally first time realize QSL state on SSL. Our studies open up a new route in searching for QSLs in frustrated quantum magnets.

**Methods**

**Sample preparation and characterization.** High-quality $Pr_2Ga_2BeO_7$ single crystals were grown by using the optical floating-zone technique. The feed and seed rods for the crystal growth were prepared by solid state reactions. The stoichiometric mixtures of $Pr_6O_{11}$ (pre-dried at 1000 ºC for overnight), $SiO_2$, and BeO were ground together and pressed into 6-mm-diameter 60-mm rods under 400-atm hydrostatic pressure and then calcined in air at 1000 ºC for 20 hours and at 1400 ºC for 20 hours, and finally in argon at 1400 ºC for 20 hours with intermediate grindings. The crystal growth was carried out in argon in an IR-heated image furnace equipped with two halogen lamps and double ellipsoidal mirrors with feed and seed rods rotating in opposite directions at 20 rpm during crystal growth at a rate of 2.0 mm/hour. A standard solid-state reaction was used to synthesize polycrystalline $La_2Ga_2BeO_7$ with $La_2O_3$ substituted for $Pr_6O_{11}$. By using the X-ray Laue photograph, the crystals were cut along the crystallographic axes for experimental studies.

**SCXRD** The single crystal of $Pr_2Ga_2BeO_7$ was picked up, mounted on the Kapton, and measured using Bruker Eco Quest Single Crystal X-ray Diffractometer (SCXRD) using Mo radiation (λ = 0.71073 Å) at 300 K. Each reflection frame was exposed for 10 s and integrated with the Bruker SAINT software package using a narrow-frame algorithm. The integration of the data using a tetragonal unit cell yielded a total of 10166 reflections to a maximum θ angle of 39.60° (0.56 Å resolution), of which 996 were independent (average redundancy 10.207, completeness = 99.7%, $R_{int}$ = 5.42%). Data were corrected for absorption effects using the multiscan method (SADABS). The ratio of minimum to maximum apparent transmission was 0.398. The structure was solved and refined using the Bruker SHELXTL Software Package. The refinement results are listed in Supplementary Information.

**Thermal conductivity measurements.** Thermal conductivity was measured by using a "one heater, two thermometers" technique in a $^3$He/$^4$He dilution refrigerator at 70 mK – 1 K and in a $^3$He refrigerator at 300 mK – 30 K, equipped with a 14 T superconducting magnet[11,12]. Two samples (A and B) were cut precisely along the crystallographic axes with the longest dimensions along the *a* or the *c* axis for the $\kappa_a$ or $\kappa_c$ measurements, respectively. The sizes of samples A and B are 2.11 × 0.53 × 0.182 and 2.13 × 0.62 ×0.159 mm$^3$, respectively. The magnetic fields were applied along either the *a* or *c* axis. Gold paint was used to make four contacts on each sample. The $RuO_2$ chip resistors were used as heaters and thermometers and are connected to the gold contacts by using gold wires and silver epoxy.

**DC magnetization measurements.** The DC susceptibility was measured using a Quantum Design SQUID-VSM, equipped with a $^3$He refrigerator insert. The DC magnetization at field up to 14 T was measured using VSM equipped in a Quantum Design PPMS.

**Specific heat measurements.** The specific heat was measured with a Quantum Design physical property measurements system (PPMS), equipped with a dilution refrigerator insert or a $^3$He refrigerator. Two thin samples of Pr$_2$Ga$_2$BeO$_7$ were cut with the thickness along the *a* or the *c* axis for the *B* // *a* or *B* // *c* measurements.

**μSR measurements.** The μSR measurements on polycrystalline Pr$_2$Ga$_2$BeO$_7$ were performed on the MuSR spectrometer at ISIS Neutron and Muon Facility, STFC, Rutherford Appleton Laboratory, UK. Around 3g powder was loaded in in an Ag-packet and mounted on an Ag-plate. The spectra in zero field (ZF) were collected in the temperature range from 30 mK to 250 K with a dilution refrigerator insert and cryostat. The asymmetry vs time *A*(*t*) μSR spectra were collected from 0.1 to 32 μs and analyzed using the WiMDA software package.

The time dependence of the muon decay asymmetry through the full temperature range (0.03 – 250 K) can be well described by the function below:

$$A_{sy}(t) = A_0 + A_1 G^{k_B T} \cdot e^{-\lambda_{tail} \cdot t} + A_2 \cdot e^{-\lambda_2 \cdot t}, \tag{1}$$

where $A_0$ is the flat background term originating from the muons stopping in the sample holder, which has been deducted from the raw data. The second term containing a classical Kubo-Toyabe function with a weak exponential relaxation rate $\lambda_{tail}$ reflects the depolarization due to the muon coupled to the nuclear magnetism. The exponential rate $\lambda_2$ is the muon spin relaxation rate usually related to the dynamic internal magnetic fields.

**Neutron scattering measurements.** INS measurements were carried out on the HOrizontally Defocusing Analyzer Concurrent Data Acquisition spectrometer (HODACA) in JRR-3, Japan[44]. Two pieces of samples, ~ 1 g, were fixed on an aluminum sheet and co-aligned in the (0KL) scattering plane. The measurements were carried out at 0.7 K with a fixed final neutron energy, Ef = 3.636 meV (energy resolution of about 0.1 – 0.2 meV). The data for different rotations were combined and analyzed with the ASYURA software package.

**CEF and DMRG calculations.** The CEF analysis is based on the point-charge model[45,46], in which the ligands surrounding the Pr$^{3+}$ magnetic ions are treated as point charges and the crystalline electrostatic potential induced can be described, in the spirit of multipolar expansion, by the following Hamiltonian

$$H_{CEF} = \sum_{n,m} B_n^m \hat{O}_n^m, \tag{2}$$

where $\hat{O}_n^m$ are Steven's operators[45] and $B_n^m$ are CEF coefficients. The CEF coefficients are calculated by using the PyCrystalField package[47], and in the calculation only the nearest eight $O^{2-}$ ligands are taken into account.

The magnetism of the compound is understood by an effective XXZ SSM (See Supplemental Information), for which the ground state is calculated by using the DMRG method[48]. We set the next nearest neighbor exchange coupling $J_2$ as the energy unit and take $J_1/J_2 = 0.685$, where the corresponding Heisenberg model has a PVBS ground state. We then examine the ground states of the XXZ model by varying the spin anisotropy $\Delta$. The calculation is performed on a $W \times L$ cylinder with width $W = 6$ and length $L = 30$. The bond dimension $D$ kept is equal to 512 and the truncation errors remain to be below $10^{-6}$.

## Data availability

The data that support the findings of this study are available from the corresponding author upon reasonable request.

## References

1. Balents, L. Spin liquids in frustrated magnets. *Nature* **464**, 199-208 (2010).
2. Savary, L. & Balents, L. Quantum spin liquids: a review. *Rep. Prog. Phys*. **80**, 016502 (2017).
3. Zhou, Y., Kanoda, K. & Ng, T.-K. Quantum spin liquid states. *Rev. Mod. Phys.* **89**, 025003 (2017).
4. Broholm, C. *et al.* Quantum spin liquids. *Science* **367**, eaay0668 (2020).
5. Chen, L. E., Schaffer, R., Sorn, S. & Kim, Y. B. Fermionic spin liquid analysis of the paramagnetic state in volborthite. *Phys. Rev. B* **96**, 165117 (2017).
6. Freire, H. Controlled calculation of the thermal conductivity for a spinon Fermi surface coupled to a U(1) gauge field. *Ann. Phys*. **349**, 357-365 (2014).
7. Faddeev, L. D. & Takhtajan, L. A. What is the spin of a spin wave? *Phys. Rev. A* **85**, 375-377 (1981).
8. Lee, S.-S., Lee, P. A. & Senthil, T. Amperean Pairing Instability in the U(1) Spin Liquid State with Fermi Surface and Application to κ-(BEDT-TTF)$_2$Cu$_2$CN$_3$. *Phys. Rev. Lett.* **98**, 067006 (2007).

9. Yamashita, M., Shibauchi, T. & Matsuda, Y. Thermal-Transport Studies on Two-Dimensional Quantum Spin Liquids, *Chem. Phys. Chem.* **13**, 74-78 (2012).

10. Yamashita, M. *et al.* Highly mobile gapless excitations in a two-dimensional candidate quantum spin liquid. *Science* **328**, 1246-1248 (2010).

11. Murayama, H. *et al.* Effect of quenched disorder on the quantum spin liquid state of the triangular-lattice antiferromagnet 1*T*-TaS$_2$. *Phys. Rev. Res.* **2**, 013099 (2020).

12. Hong, X. *et al.* Spinon Heat Transport in the Three-Dimensional Quantum Magnet PbCuTe$_2$O$_6$, *Phys. Rev. Lett.* **131**, 256701 (2023).

13. Rao, X. *et al.* Survival of itinerant excitations and quantum spin state transitions in YbMgGaO$_4$ with chemical disorder, *Nat. Commun.* **12**, 4949 (2021).

14. Li, N. *et al.* Possible itinerant excitations and quantum spin state transitions in the effective spin-1/2 triangular-lattice antiferromagnet Na$_2$BaCo(PO$_4$)$_2$. *Nat. Commun.* **11**, 4216 (2020).

15. Bourgeois-Hope, P. *et al.* Thermal conductivity of the quantum spin liquid candidate EtMe$_3$Sb[Pd(dmit)$_2$]$_2$: No evidence of mobile gapless excitations. *Phys. Rev. X* **9**, 041051 (2019).

16. Ni, J. M. *et al.* Absence of magnetic thermal conductivity in the quantum spin liquid candidate EtMe$_3$Sb[Pd(dmit)$_2$]$_2$ – revisited. *Phys. Rev. Lett.* **123**, 247204 (2019).

17. Yu, Y. J. *et al.* Heat transport study of the spin liquid candidate 1*T*-TaS$_2$. *Phys. Rev. B* **96**, 081111(R) (2017).

18. Xu, Y. *et al.* Absence of Magnetic Thermal Conductivity in the Quantum Spin-Liquid Candidate YbMgGaO$_4$. *Phys. Rev. Lett.* **117**, 267202 (2016).

19. Shastry, B. S. & Sutherland, B. Exact ground state of a quantum mechanical antiferromagnet. *Physica B* (Amsterdam) **108**, 1069-1070 (1981).

20. Kageyama, H. *et al.* Exact Dimer Ground State and Quantized Magnetization Plateaus in the Two-Dimensional Spin System SrCu$_2$(BO$_3$)$_2$. *Phys. Rev. Lett*. **82**, 3168 (1999).

21. Kodama, K. *et al.* Magnetic Superstructure in the Two-Dimensional Quantum Antiferromagnet SrCu$_2$(BO$_3$)$_2$. *Science* **298**, 395-399 (2002).

22. Huang, H. X., Chen, Y., Gao, Y. & Yang, G. H. Unconventional superconducting states on doped Shastry–Sutherland lattice. *Physica C* **525-526**, 1-4 (2016).

23. Cui, Y. *et al.* Proximate deconfined quantum critical point in SrCu$_2$(BO$_3$)$_2$. *Science* **380**, 1179-1184 (2023).

24. Jiménez, J. L. *et al.* A quantum magnetic analogue to the critical point of water. *Nature* **592**, 370-375 (2021).

25. Zayed, M. E. *et al.* 4-spin plaquette singlet state in the Shastry–Sutherland compound $SrCu_2(BO_3)_2$. *Nat. Phys.* **13**, 962-966 (2017).

26. McClarty, P. A. *et al.* Topological triplon modes and bound states in a Shastry–Sutherland magnet. Nat. Phys. **13**, 736-742 (2017).

27. Sun, X. F. & Ando, Y. Comment on "Low-temperature phonon thermal conductivity of single-crystalline $Nd_2CuO_4$: Effects of sample size and surface roughness", *Phys. Rev. Lett.* **79**, 176501 (2009).

28. Zeng, Z. *et al.* Possible Dirac quantum spin liquid in the kagome quantum antiferromagnet $YCu_3(OH)_6Br_2[Br_x(OH)_{1-x}]$. *Phys. Rev. B* **105**, L121109 (2022).

29. Xu, S. *et al.* Realization of U(1) Dirac Quantum Spin Liquid in $YbZn_2GaO_5$, arXiv:2305.20040 (2023).

30. Bu, H. *et al.* Gapless triangular-lattice spin-liquid candidate $PrZnAl_{11}O_{19}$. *Phys. Rev. B* **106**, 134428 (2022).

31. Ma, Z. *et al.* Possible gapless quantum spin liquid behavior in the triangular-lattice Ising antiferromagnet $PrMgAl_{11}O_{19}$. *Phys. Rev. B* **109**, 165143 (2024).

32. Kundu, S. *et al.* Gapless Quantum Spin Liquid in the Triangular System $Sr_3CuSb_2O_9$. *Phys. Rev. Lett*. **125**, 267202 (2020).

33. Ran, Y., Hermele, M., Lee, P. A. & Wen, X.-G. Projected-wave-function study of the spin-1/2 Heisenberg model on the Kagomé lattice, *Phys. Rev. Lett.* **98**, 117205 (2007).

34. Hu, S., Zhu, W., Eggert, S. & He, Y.-C. Dirac Spin Liquid on the Spin-1/2 Triangular Heisenberg Antiferromagnet. *Phys. Rev. Lett.* **123**, 207023 (2019).

35. Clark, L. *et al.* Two-dimensional spin liquid behaviour in the triangular-honeycomb antiferromagnet $TbInO_3$. *Nat. Phys*. **15**, 262-268 (2019).

36. Li, Y. *et al.* Muon Spin Relaxation Evidence for the U(1) Quantum Spin-Liquid Ground State in the Triangular Antiferromagnet $YbMgGaO_4$. *Phys. Rev. Lett.* **117**, 097201 (2016).

37. Ding, L. *et al.* Gapless spin-liquid state in the structurally disorder-free triangular antiferromagnet $NaYbO_2$. *Phys. Rev. B* **100**, 144432 (2019).

38. Arh, T. *et al.* The Ising triangular-lattice antiferromagnet neodymium heptatantalate as a quantum spin liquid candidate. *Nat. Mater.* **21**, 416-422 (2022).

39. Balz, C. *et al.* Physical realization of a quantum spin liquid based on a complex frustration mechanism. *Nat. Phys.* **12**, 942-949 (2016).

40. Zhu, Z. H. et al. Fluctuating magnetic droplets immersed in a sea of quantum spin liquid. *The Innovation* **4**, 100459 (2023).

41. Xi, N., Chen, H., Xie, Z. Y. & Yu, R. Plaquette valence bond solid to antiferromagnet transition and deconfined quantum critical point of the Shastry-Sutherland model. *Phys. Rev. B* **107**, L220408 (2023).

42. Yang, J., Sandvik, A. W. & Wang, L. Quantum criticality and spin liquid phase in the Shastry-Sutherland model. *Phys. Rev. B* **105**, L060409 (2022).

43. Kikuchi, H., Asai, S., Sato, T. J., Nakajima, T., Harriger, L., Zaliznyak, I., Masuda, T. A new inelastic neutron spectrometer HODACA. arXiv:2310.11463 (2023).

44. Hussey, N. E. Low-energy quasiparticles in high-$T_c$ cuprates. *Adv. Phys.* **51**, 1685-1771 (2002).

45. Stevens, K. W. H. Matrix Elements and Operator Equivalents Connected with the Magnetic Properties of Rare Earth Ions, *Proc. Phys. Soc. A* **65**, 209-214 (1952).

46. Hutchings, M. T. Point-Charge Calculations of Energy Levels of Magnetic Ions in Crystalline Electric Fields, *Solid State Phys.* **16**, 227-273 (1964).

47. Scheie, A., PyCrystalField: Software for Calculation, Analysis, and Fitting of Crystal Electric Field Hamiltonians. *J. Appl. Cryst.* **54**, 356-362 (2021).

48. White, S. R. Density matrix formulation for quantum renormalization groups. *Phys. Rev. Lett.* **69**, 2863-2866 (1992).

**Acknowledgements**

We thank Z.-X. Liu for helpful discussion. This work was supported by the National Key Research and Development Program of China (Grant Nos. 2023YFA1406500 and 2022YFA1402702), the National Natural Science Foundation of China (Grant Nos. U2032213, 12334008, 12274388, 12174361, 12174441, 12104010, and 12104011) and the Nature Science Foundation of Anhui Province (Grant Nos. 1908085MA09 and 2108085QA22). The work at the University of Tennessee was supported by the NSF with Grant No. NSF-DMR-2003117. H.W. and W.X. are supported by the U.S. DOE-BES under Contract No. DE-SC0023648.

**Author contributions**

N.L. and X.F.S. performed thermal conductivity, magnetization and specific heat measurements and analyzed the data with help from Y.Y.W., H.L., Q.J.L. and X.Z. A.B. and H.D.Z. grew the crystals and measured magnetization. H.W. and W.X. performed SCXRD measurement and refinement. M.F.S., J.M. and R.Z. measured the specific heat. M.F.S., J.M. and P.J.B. measured the $\mu$SR. M.F.S., J.M., H.K. and T.M. measured the neutron scattering. G.D., C.L. and R.Y. performed theoretical analysis. N.L., X.F.S., H.D.Z., R.Y., and J.M. wrote the paper with input from all other co-authors.

**Competing financial interests**

The authors declare no competing interests.

**Additional Information**

**Correspondence and requests for materials** should be addressed to X.F.S., H.D.Z., R.Y. or J.M.

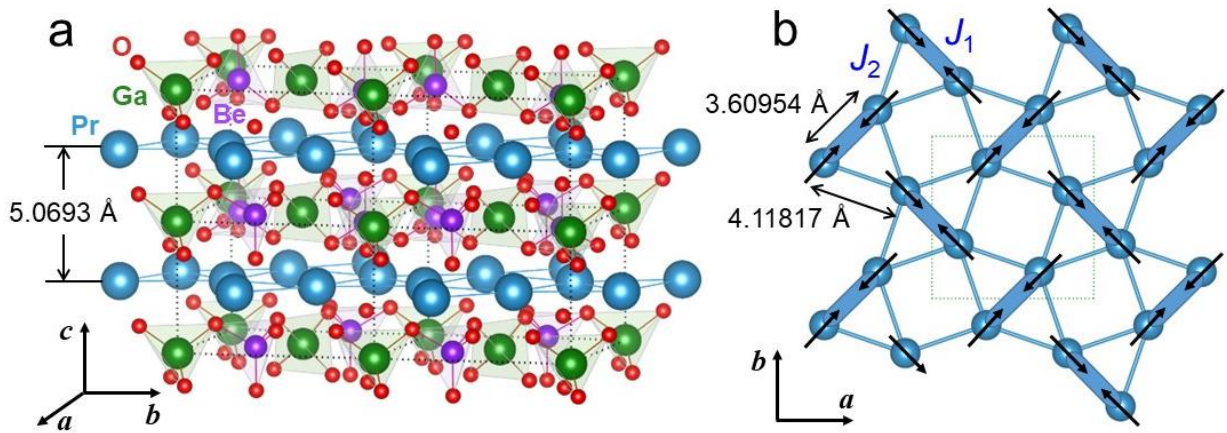

**Figure 1 Structure of Pr$_2$Ga$_2$BeO$_7$ single crystal. a,** Schematic illustration of the Shastry-Sutherland layer in Pr$_2$Ga$_2$BeO$_7$. The dimer bond consists of a pair of spins, as represented by the arrows. The $J_1$ and $J_2$ indicate the intradimer and interdimer exchange, respectively.

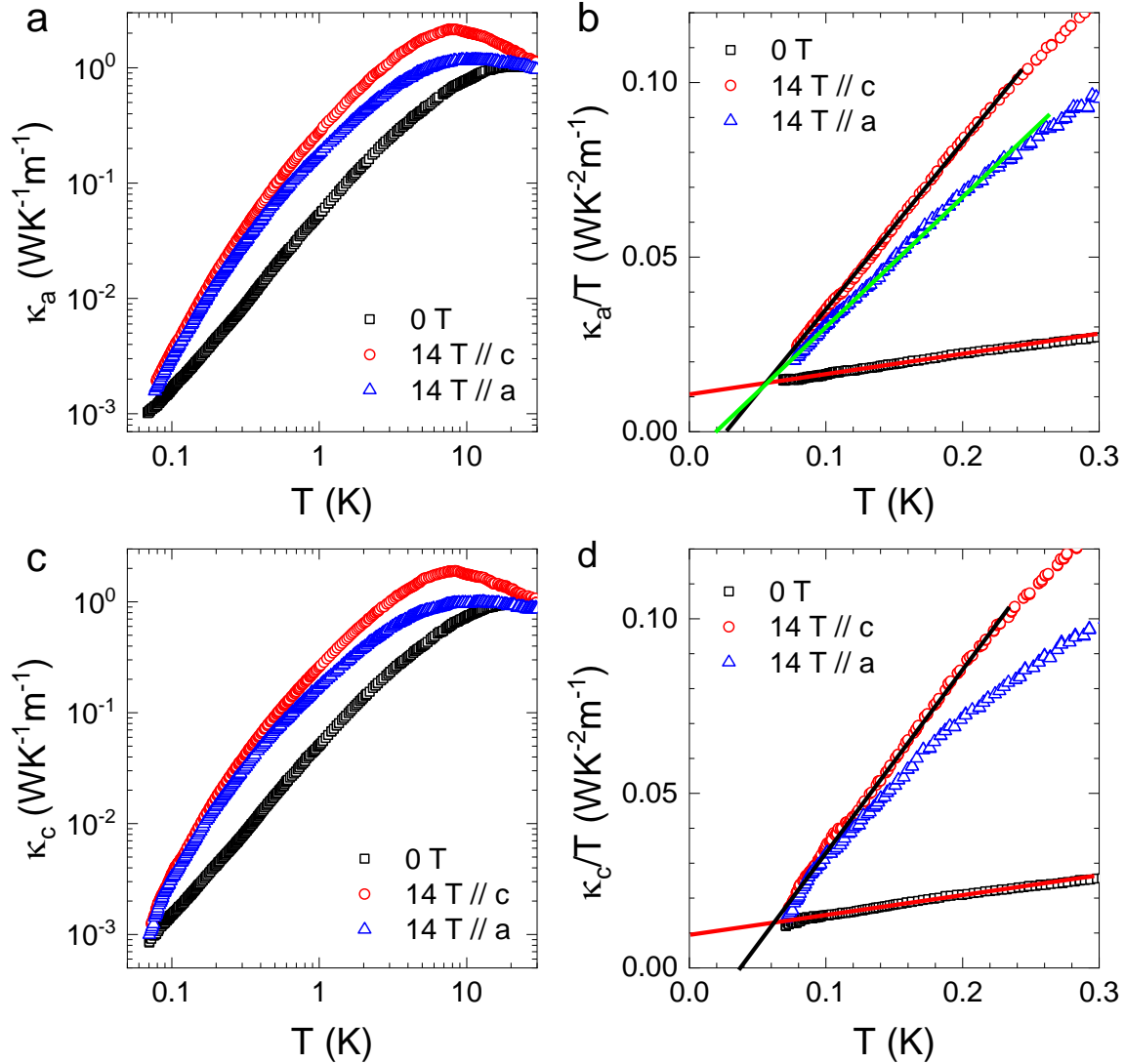

**Figure 2 Low-temperature thermal conductivity of $Pr_2Ga_2BeO_7$ single crystals with heat current along the *a* or *c* axis. a,** Temperature dependence of $\kappa_a$ at zero field and 14 T magnetic field applied along either the *c* or *a* axis. **b,** $\kappa_a/T$ plotted as a function $T$ at $T < 0.3$ K. The solid lines are the fittings to the low-temperature data by using the formula $\kappa/T = a + bT$, where $a = \kappa_0/T$. **c,** Temperature dependence of $\kappa_c$ at zero field and 14 T magnetic field applied either along or perpendicular to the *c* axis. **d,** $\kappa_c/T$ plotted as a function $T$ at $T < 0.3$ K. The solid lines are the fittings to the low-temperature data by using the formula $\kappa/T = a + bT$, where $a = \kappa_0/T$.

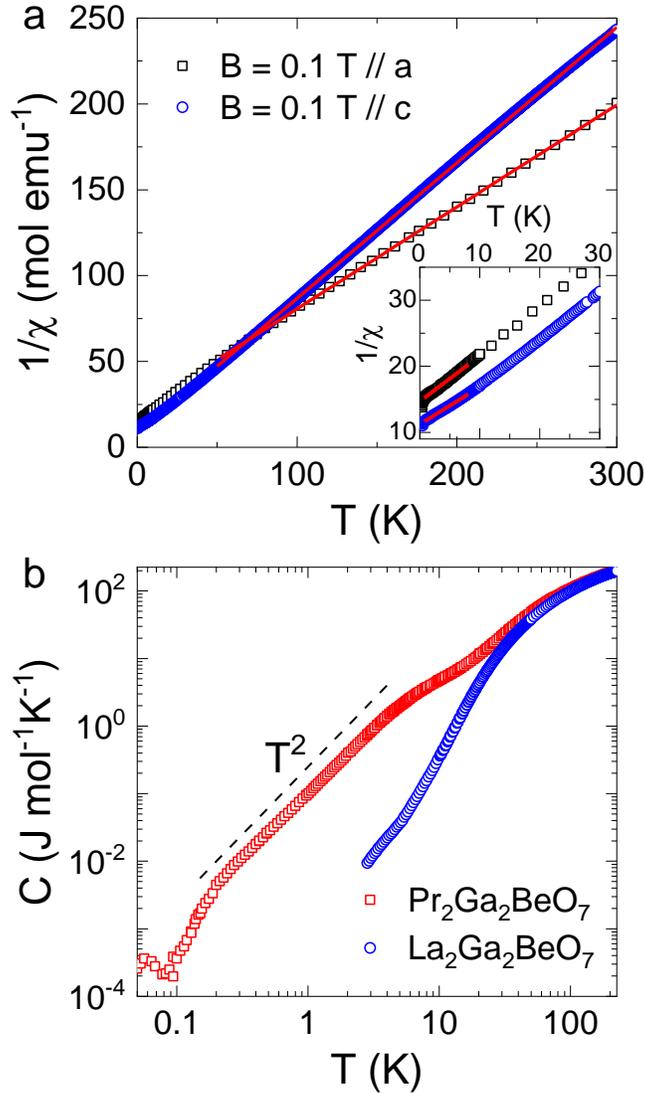

**Figure 3  Magnetic susceptibility and specific heat of Pr$_2$Ga$_2$BeO$_7$ single crystal. a,** Temperature dependence of the inverse magnetic susceptibility with magnetic field ($B$ = 0.1 T) along the $c$ or $a$ axis. The inset enlarges the low temperature range and the red lines indicate the CW fitting at the selected temperature range. **b,** Specific heat of Pr$_2$Ga$_2$BeO$_7$ single crystal. Also shown are the data for nonmagnetic La$_2$Ga$_2$BeO$_7$ polycrystal, which can be taken as the phononic term of specific heat in Pr$_2$Ga$_2$BeO$_7$. Note that at low temperatures ($T$ < 4 K) the phononic specific heat is negligible in Pr$_2$Ga$_2$BeO$_7$, since it further decays in a $T^3$ way. The dashed line indicates the $T^2$ dependence at very low temperatures.

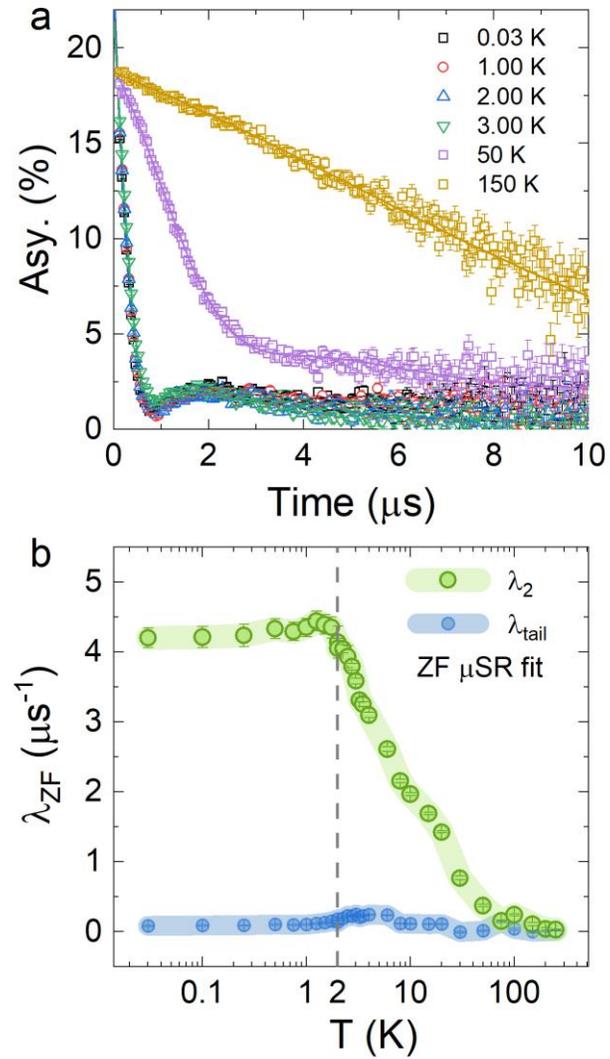

**Figure 4 Low-temperature $\mu$SR data of Pr$_2$Ga$_2$BeO$_7$. a,** Zero-field muon decay asymmetry data measured on the $\mu$SR spectrometer at various temperatures, with solid lines showing fits of Equation (1). **b,** Muon spin relaxation rate $\lambda$ as a function of temperature.

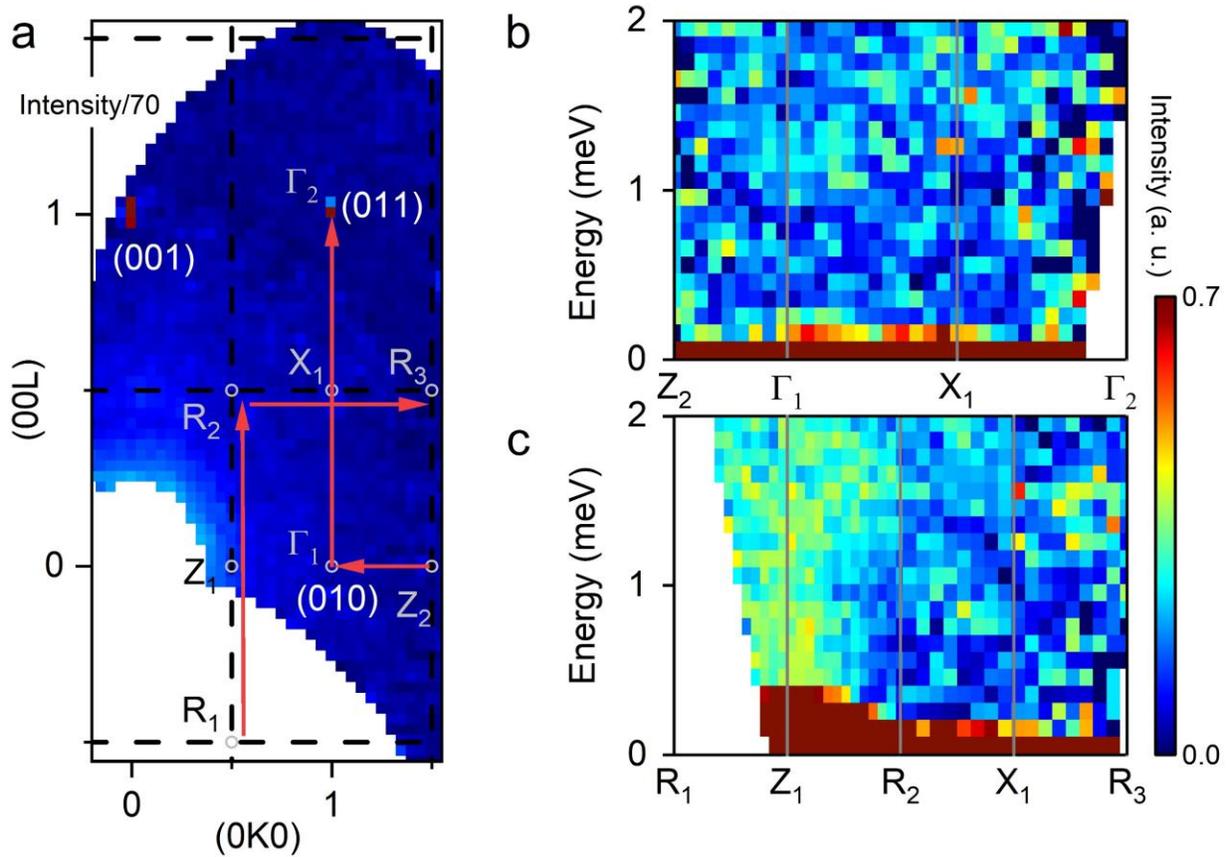

**Figure 5 Single-crystal inelastic neutron scattering results for Pr$_2$Ga$_2$BeO$_7$. a,** Elastic magnetic scattering in the 0KL plane, the black dashed lines represent the Brillouin zone boundaries. The grey hollow circles represent high symmetry points R$_1$, Z$_1$, R$_2$, X$_1$, R$_3$, Z$_2$, Γ$_1$ and Γ$_2$ marked in (a). **b** and **c,** Spin-excitation spectra along high symmetry momentum directions R$_1$-Z$_1$-R$_2$-X$_1$-R$_3$ and Z$_2$-Γ$_1$-X$_1$-Γ$_2$, respectively.

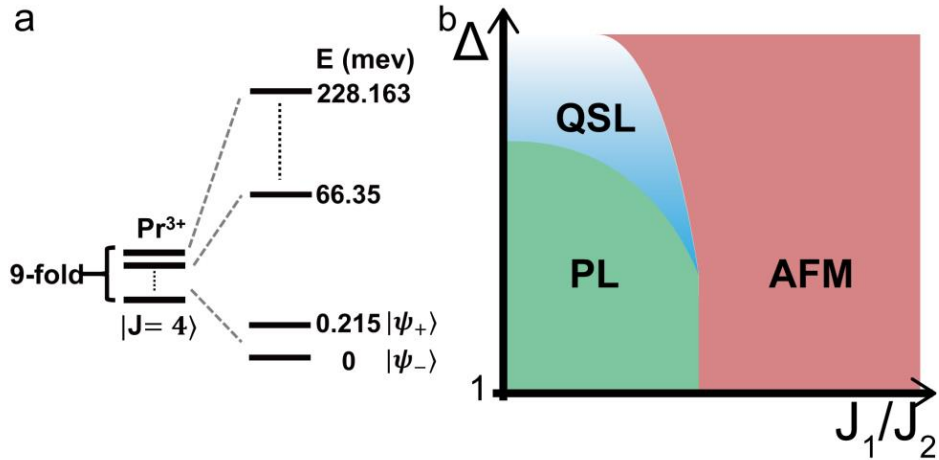

**Figure 6 Ground state of Pr-based SSM**. **a,** Sketch of the single-ion crystal field levels obtained from the point-charge model, the ground-state manifold contains a quasi-degenerate non-Kramers doublet which are well separated from other high-energy levels. **b,** Sketched phase diagram of the XXZ SSM. The plaquette (PL) to AFM transition is close to a DQCP in the Heisenberg case ($\Delta = 1$). A gapless QSL is stabilized between the PL and AFM ground states by increasing the spin anisotropy $\Delta$.

# Supplementary Information for

## "Spinons in a new Shastry-Sutherland lattice magnet $Pr_2Ga_2BeO_7$"

N. Li[1,14], A. Brassington[2,14], M. F. Shu[3,14], Y. Y. Wang[1], H. Liang[1], Q. J. Li[4], X. Zhao[5], P. J. Baker[6], H. Kikuchi[7], T. Masuda[7,8], G. Duan[9], C. Liu[10], H. Wang[11], W. Xie[11], R. Zhong[12], J. Ma[3]★, R. Yu[9,13]★, H. D. Zhou[2]★, and X. F. Sun[1]★

[1]Anhui Key Laboratory of Magnetic Functional Materials and Devices, Institutes of Physical Science and Information Technology, Anhui University, Hefei, Anhui 230601, People's Republic of China

[2]Department of Physics and Astronomy, University of Tennessee, Knoxville, Tennessee 37996-1200, USA

[3]Key Laboratory of Artificial Structures and Quantum Control, Shenyang National Laboratory for Materials Science, School of Physics and Astronomy, Shanghai Jiao Tong University, Shanghai 200240, People's Republic of China

[4]School of Physics and Optoelectronics, Anhui University, Hefei, Anhui 230061, People's Republic of Chin

[5]School of Physics Sciences, University of Science and Technology of China, Hefei, Anhui 230026, People's Republic of China

[6]ISIS Facility, Rutherford Appleton Laboratory, Chilton, Didcot OX11 0QX, United Kingdom

[7]Institute for Solid State Physics, The University of Tokyo, Chiba 277-8581, Japan

[8]Institute of Materials Structure Science, High Energy Accelerator Research Organization, Ibaraki 305-0801, Japan

[9]Department of Physics and Beijing Key Laboratory of Opto-electronic Functional Materials and Micro-nano Devices, Renmin University of China, Beijing 100872, People's Republic of China

[10]School of Engineering, Dali University, Dali, Yunnan 671003, People's Republic of China

[11]Department of Chemistry, Michigan State University, East Lansing, Michigan 48824, USA

[12]Tsung-Dao Lee Institute, Shanghai Jiao Tong University, Shanghai 201210, People's Republic of China

[13]Key Laboratory of Quantum State Construction and Manipulation (Ministry of Education), Renmin University of China, Beijing, 100872, People's Republic of China

[14]These authors contributed equally: N. Li, A. Brassington, M. F. Shu

★email: jma3@sjtu.edu.cn; rong.yu@ruc.edu.cn; hzhou10@utk.edu; xfsun@ahu.edu.cn

**Supplementary Table SI.** Materials with other supporting evidence for QSL state but reported zero $\kappa_0/T$ term.

| Material | Structure | Other signatures for QSL |
|---|---|---|
| $\kappa$-(BEDT-TTF)$_2$Cu$_2$(CN)$_3$ | triangular | linear-$T$ dependent specific heat[S1,S2] |
| ZnCu$_3$(OH)$_6$Cl$_2$ | kagome | power law of specific heat, continuum mode[S3-S5] |
| YCu$_3$(OH)$_{6.5}$Br$_{2.5}$ | kagome | $T^2$ dependent specific heat, continuum mode[S6-S8] |
| NaYbSe$_2$ | triangular | linear-$T$ dependent specific heat, continuum mode, plateau of muon spin relaxation rate[S9-S11] |
| Ca$_{10}$Cr$_7$O$_{28}$ | bilayer kagome | continuum mode, plateau of muon spin relaxation rate, linear-$T$ dependent specific heat[S12-14] |

*Crystal structure:*

The detailed crystallographic parameters and atomic coordinates obtained from the refinement of the single crystal X-ray diffraction data are listed in Tables SII and SIII, respectively.

**Supplementary Table SII.** The crystal structure and refinement for $Pr_2Ga_2BeO_7$ at 300 K.

| Chemical Formula | $Pr_2Ga_2BeO_7$ |
|---|---|
| Formula weight | 542.27 g/mol |
| Space Group | $P\text{-}42_1m$ |
| Unit cell dimensions | $a$ = 7.7870(14) Å<br>$c$ = 5.0693(14) Å |
| Volume | 307.39(14) Å$^3$ |
| Density (calculated) | 5.859 g/cm$^3$ |
| Extinction coefficient | 0.1270(110) |
| Absorption coefficient | 24.224 mm$^{-1}$ |
| F(000) | 480 |
| $2\theta$ range | 7.40 to 79.20° |
| Total Reflections | 10166 |
| Independent reflections | 996 [$R_{int}$ = 0.0542] |
| Refinement method | Full-matrix least-squares on $F^2$ |
| Absolute structure parameter | 0.14(8) |
| Data / restraints / parameters | 996 / 0 / 37 |
| Final R indices | $R_1$ (I>2σ(I)) = 0.0393; $wR_2$ (I>2σ(I)) = 0.1142<br>$R_1$ (all) = 0.0398; $wR_2$ (all) = 0.1145 |
| Largest diff. peak and hole | +8.806 e/Å$^{-3}$ and -6.127 e/Å$^{-3}$ |
| R.M.S. deviation from mean | 2.274 e/Å$^{-3}$ |
| Goodness-of-fit on $F^2$ | 1.472 |

**Supplementary Table SIII.** Atomic coordinates and equivalent isotropic atomic displacement parameters (Å$^2$). ($U_{eq}$ is defined as one third of the trace of the orthogonalized $U_{ij}$ tensor.)

| $Pr_2Ga_2BeO_7$ | Wyck. | $x$ | $y$ | $z$ | Occ. | $U_{eq}$ |
|---|---|---|---|---|---|---|
| Pr | 4e | 0.66388(4) | 0.16388(4) | 0.49661(13) | 1 | 0.0171(2) |
| Ga$_1$ | 2a | 0 | 0 | 0 | 1 | 0.0090(3) |
| Ga$_2$ | 4e | 0.14161(16) | 0.64161(16) | 0.0380(4) | 0.493 | 0.0062(4) |
| Be | 4e | 0.14161(16) | 0.64161(16) | 0.0380(4) | 0.507 | 0.0062(4) |
| O$_1$ | 2c | 0 | ½ | 0.189(2) | 1 | 0.0122(14) |
| O$_2$ | 4e | 0.1395(7) | 0.6395(7) | 0.7073(15) | 1 | 0.0153(11) |
| O$_3$ | 8f | 0.3286(7) | 0.5821(7) | 0.2135(10) | 1 | 0.0136(8) |

The X-ray powder diffraction data and the refinement of ground single crystal of $Pr_2Ga_2BeO_7$ are shown in Fig. S1(a). The refinement results from the powder X-ray diffraction data are consistent with the single crystal X-ray diffraction results. Figure S1(b) shows the photo of $Pr_2Ga_2BeO_7$ single crystal, which was grown by using the floating-zone technique. Figure S1(c) shows the X-ray Laue back reflection of the *ab* plane, in which the sharp reflection points indicate good crystallinity of the sample.

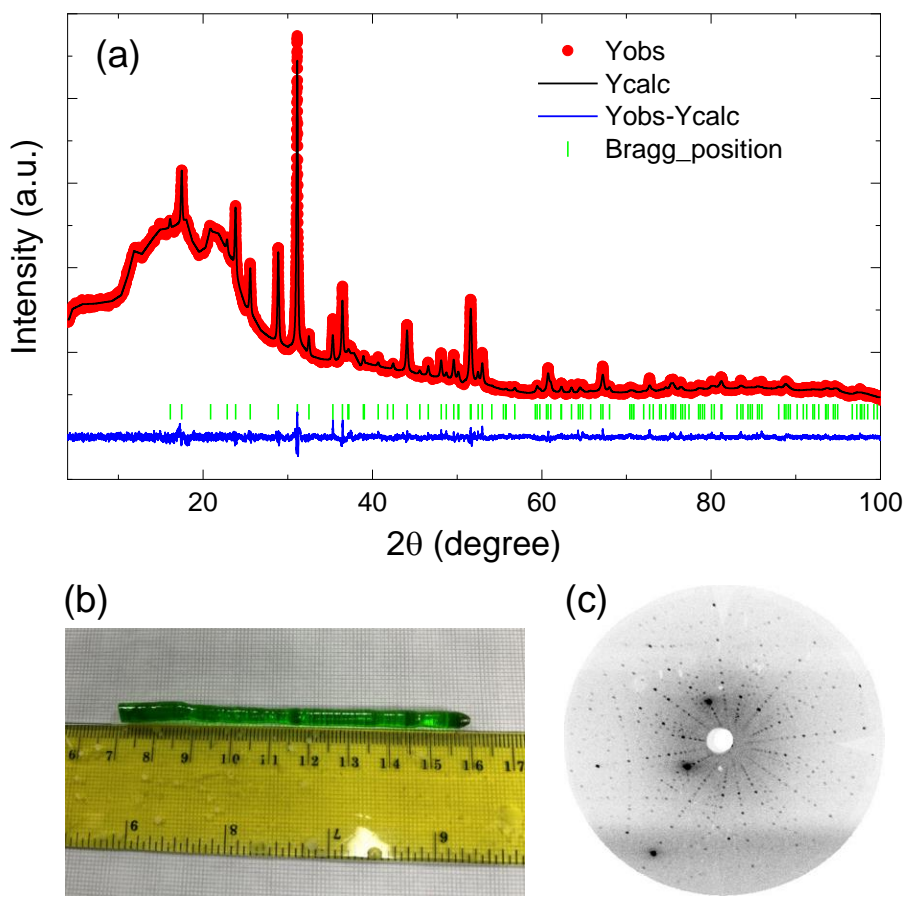

**Supplementary Figure S1** (a) The XRD data and the refinement pattern of $Pr_2Ga_2BeO_7$. (b) The photo of the as-grown $Pr_2Ga_2BeO_7$ single crystal. (c) The X-ray Laue back reflection of the *ab* plane.

*Magnetization:*

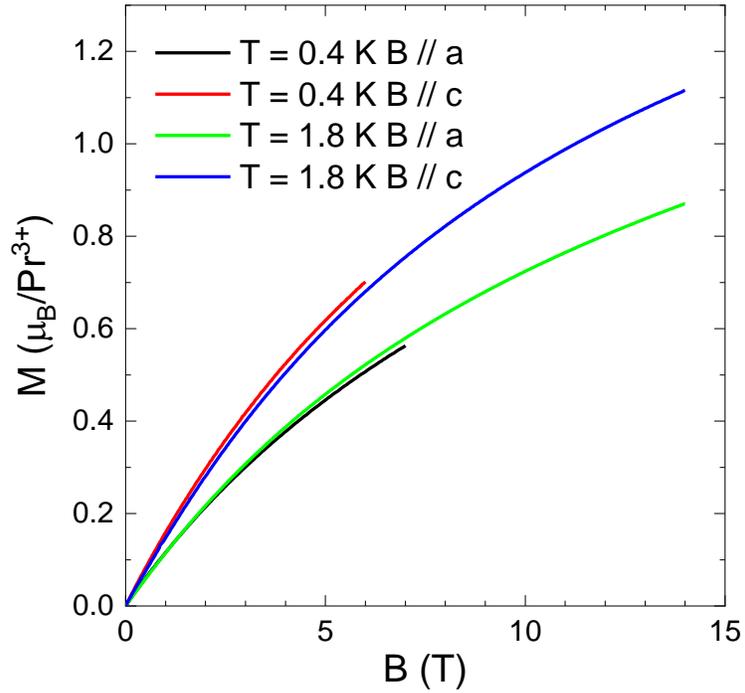

**Supplementary Figure S2** Magnetic field dependence of the magnetization at $T = 0.4$ and 1.8 K for magnetic field along the *c* or *a* axis.

Figure S2 shows the isothermal magnetization $M(B)$ curves of $Pr_2Ga_2BeO_7$ single crystal at $T = 0.4$ and 1.8 K with external magnetic field along the *c* or *a* axis. As we can see, the magnetization exhibits a weak temperature dependence and a non-linear magnetic field response. The magnetization does not achieve the saturated value ($g_J J \mu_B$) of free $Pr^{3+}$ ions for field up to 14 T, which may be related to the low-energy CEF effects. Moreover, no field-induced transitions are detected for all curves and the magnetization along the *c* axis is larger, which means a weak spin anisotropy.

*Specific heat:*

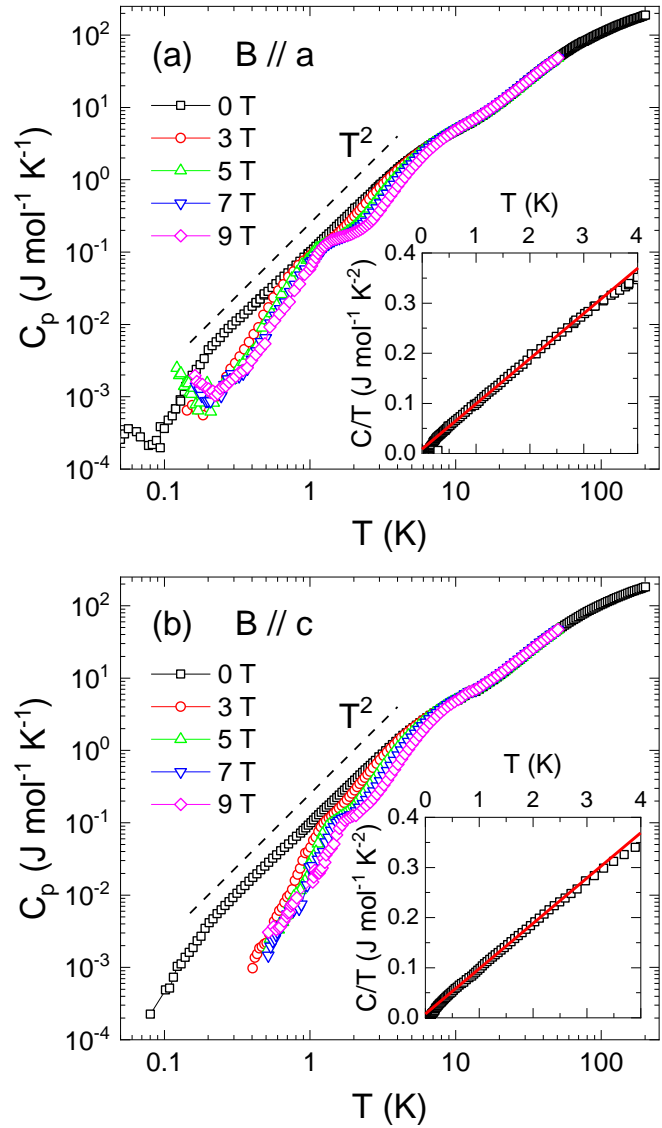

**Supplementary Figure S3** Low-temperature specific heat of $Pr_2Ga_2BeO_7$ single crystals under different magnetic fields along the *a* axis (a) and *c* axis (b). The dashed lines indicate the $T^2$ dependence. The insets show the *C/T vs T* plots for the low-temperature data of these two samples at zero field. The small but finite zero-temperature intercepts of the linear fitting (red lines) confirm the existence of residual density of states.

Figure S3 shows the low-temperature specific heat of $Pr_2Ga_2BeO_7$ single crystal at various magnetic fields along the *a* axis. The external magnetic field can significantly suppress the specific heat below 10 K, indicating the predominately magnetic contribution at low temperatures.

Moreover, a field-induced anomaly is observed around 1 K and it becomes more significant and shifts to higher temperature with increasing magnetic field. The similar behaviors were observed for $B // c$, as shown in Fig. S3(b). It is notable that the low-temperature specific heat at zero field, which is mainly the magnetic contribution, displays a $T^2$ dependence for $0.2 < T < 4$ K, as shown in Fig. S3(a).

*Thermal conductivity:*

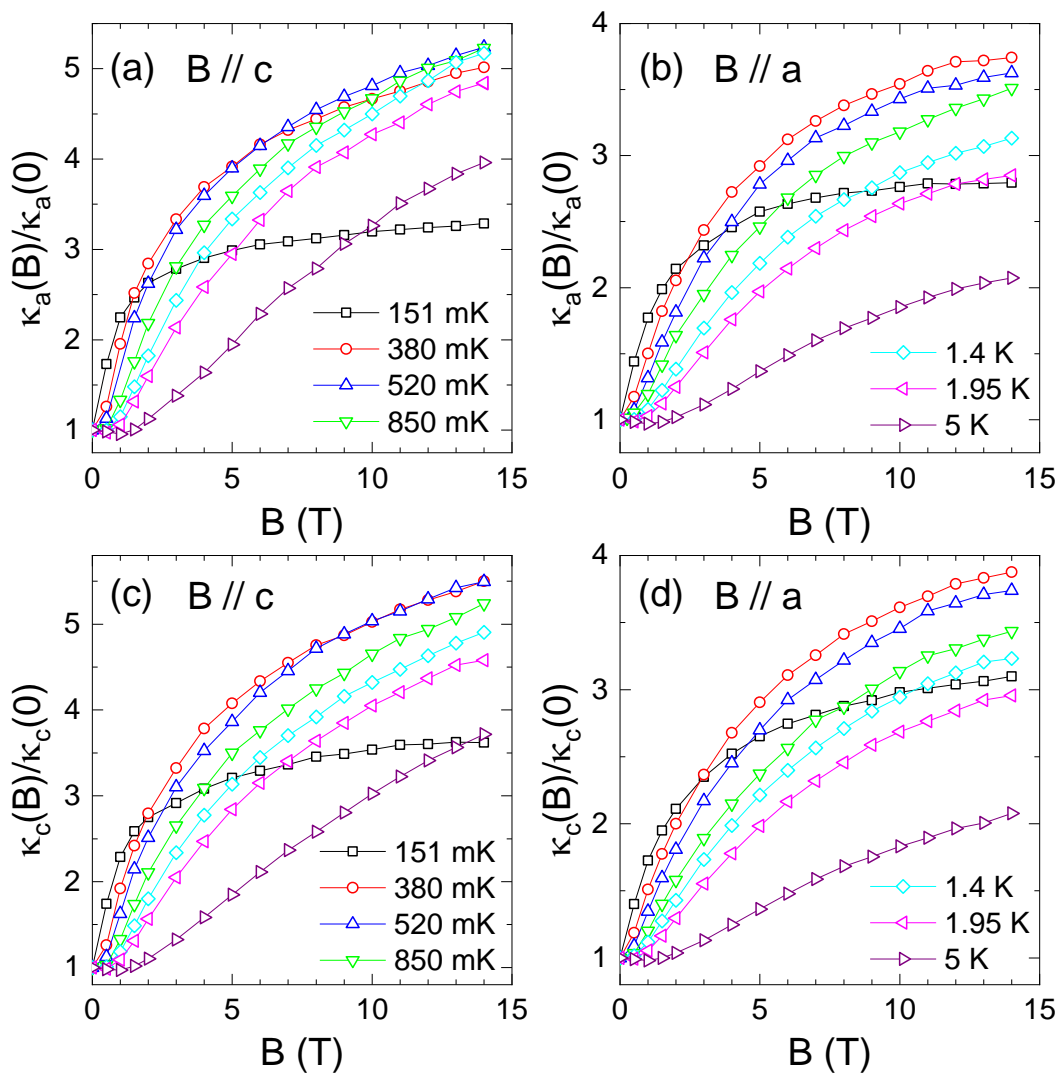

**Supplementary Figure S4** The field dependence of $\kappa_a$ (sample A) and $\kappa_c$ (sample B) at selected temperatures with external magnetic field along the *c* and *a* axis, respectively.

The magnetic field dependence of $\kappa_a$ (sample A) is shown in Figs. S4(a) and S4(b) for the external magnetic field along the *c* and *a* axis, respectively. As one can see that the $\kappa_a$ increases monotonically with increasing magnetic field, and there is no field-induced transition. At $T = 151$ mK, the $\kappa_a(B)$ curve displays weak field dependence at high magnetic fields, where the spins are polarized and the magnetic scattering effect on phonons is nearly completely suppressed. Moreover, the saturation field shifts to higher magnetic field with increasing temperature and the slight difference between $B // a$ and $B // c$ indicates a weak anisotropy in $Pr_2Ga_2BeO_7$. The behavior of $\kappa_c(B)/\kappa_c(0)$ curves (sample B) is similar to that of $\kappa_a$, as shown in Figs. S4(c) and S4(d).

We also measured the thermal conductivity of another two samples (C and D) to check the reproducibility of the experimental results. These two samples have the sizes of $2.69 \times 0.64 \times 0.164$ mm$^3$ and $2.20 \times 0.60 \times 0.151$ mm$^3$, respectively, with the longest dimension along the $a$ axis. That is, the heat current was applied along the $a$ axis for them. As shown in Figs. S5 and S6, both of samples C and D show comparable thermal conductivity data to those of the $\kappa_a$ (sample A) shown in the main text. Particularly, these two samples also exhibit finite residual $\kappa_0/T$ of 0.0126 and 0.0140 WK$^{-2}$m$^{-1}$, respectively.

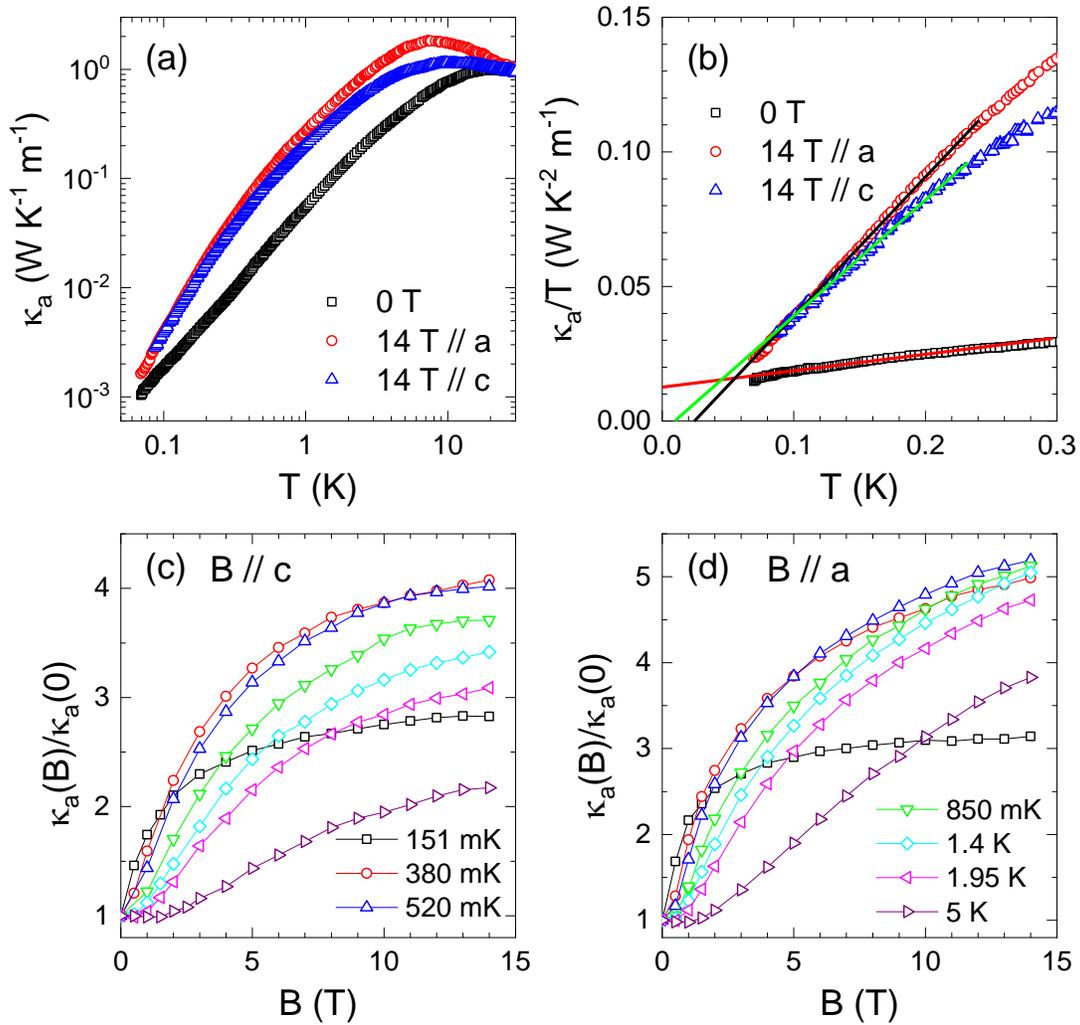

**Supplementary Figure S5** Thermal conductivity of sample C with heat current along the *a* axis. (a) Temperature dependence of $\kappa_a$ at zero field and 14 T magnetic field applied along either the *c* or *a* axis. (b) $\kappa_a/T$ plotted as a function $T$ at $T < 0.3$ K. The solid lines are the fittings to the low-temperature data by using the formula $\kappa/T = a + bT$ ($a = \kappa_0/T$). (c,d) The field dependence of $\kappa_a$ at selected temperatures with external magnetic field along the *c* and *a* axis, respectively.

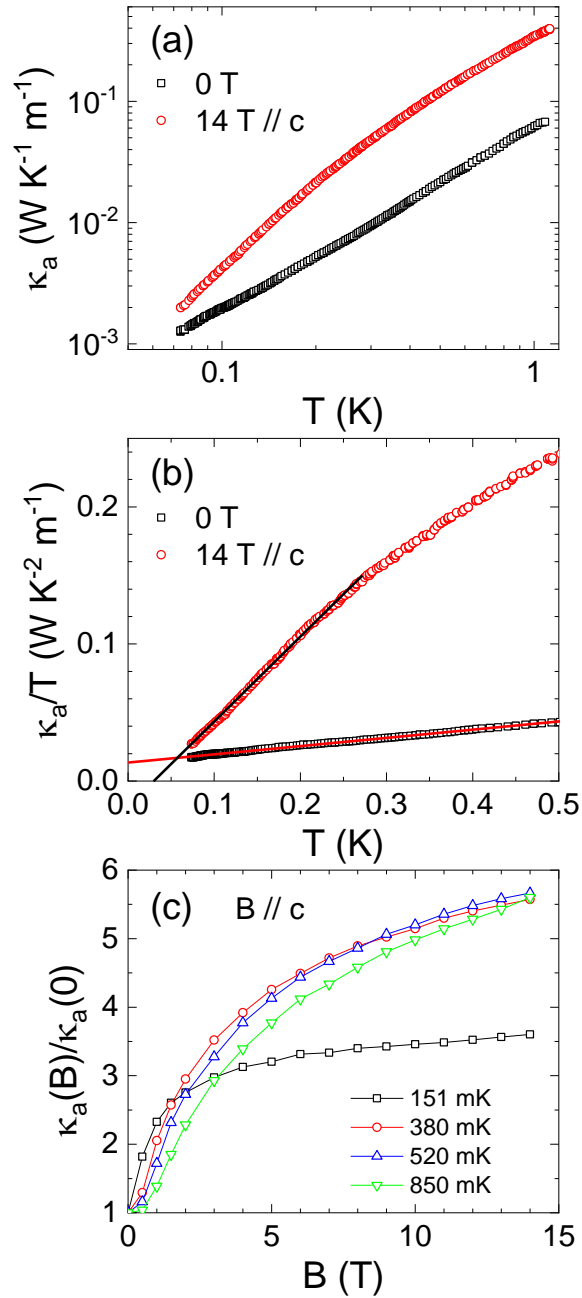

**Supplementary Figure S5** Thermal conductivity of sample D with heat current along the *a* axis. (a) Temperature dependence of $\kappa_a$ at zero field and 14 T magnetic field applied along the *c* axis. (b) $\kappa_a/T$ plotted as a function $T$ at $T < 0.5$ K. The solid lines are the fittings to the low-temperature data by using the formula $\kappa/T = a + bT$ ($a = \kappa_0/T$). (c) The field dependence of $\kappa_a$ at selected temperatures with external magnetic field along the *c* axis.

*Theoretical modeling and analysis:*

In this section, we construct an effective model to understand the magnetism of $Pr_2Ga_2BeO_7$. We start from the single-ion physics. As discussed in the main text, the non-Kramers $Pr^{3+}$ ion has a $4f^2$ electron configuration with a total orbital angular momentum $L = 5$ and a total spin angular momentum $S = 1$. $L$ and $S$ are combined to the total angular momentum $J = 4$ under strong spin-orbit coupling (SOC). In a CEF environment, the 9-fold degenerate $J = 4$ multiplet splits into various crystal field levels. In this compound, the local point group symmetry of the $Pr^{3+}$ site is very low. In each Pr layer, the only symmetry operation is the mirror plane about either the [1-10] (denoted as $x'$) or [110] (denoted as $y'$) direction, as illustrated in Fig. S7. As a consequence, there is no symmetry protected multiplets under CEF, and all crystal field levels are singlets. To check this, we performed calculations with the PyCrystalField package[S15] by assuming a point-charge CEF model. The energies and corresponding wave functions of the CEF levels in the model are given in Table SIV (in the crystallographic basis).

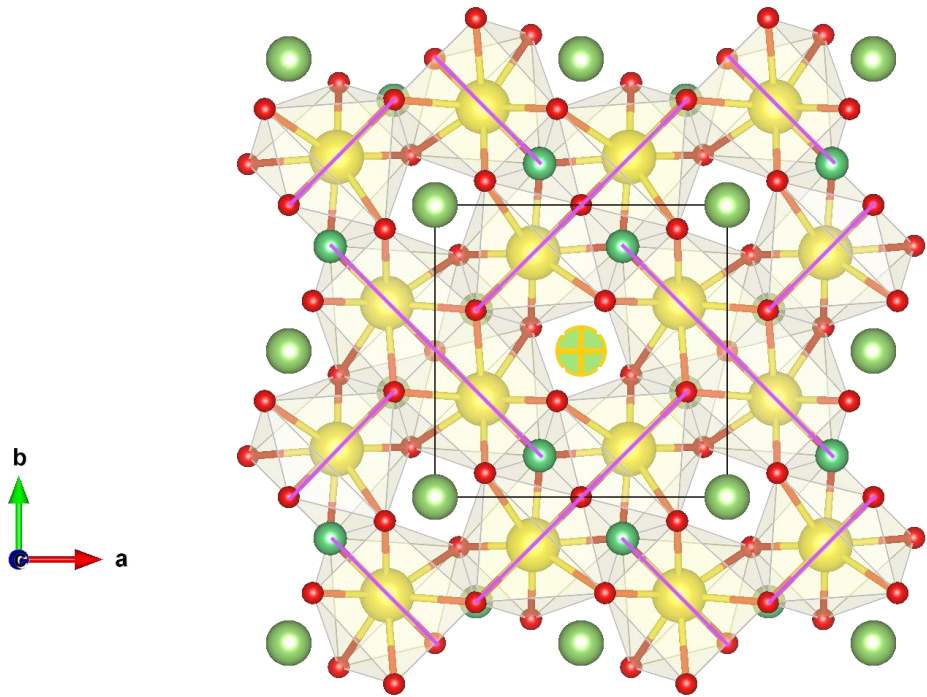

**Supplementary Figure S7** Crystal structure of $Pr_2Ga_2BeO_7$. $Pr^{3+}$ and $O^{2-}$ are denoted by yellow and red balls, respectively. Mirror planes of $Pr^{3+}$ are denoted by purple lines.

**Supplementary Table SIV.** Eigenvectors and Eigenvalues of CEF levels of $Pr^{3+}$ at [0.66388,0.16388,0.49661] (l.u.). The axes are chosen the same as the crystallographic axes. The first two low-energy levels, denoted as $|\psi_-\rangle$ and $|\psi_+\rangle$, respectively, constitute a non-Kramers doublet.

| E (meV) | $J^z=|-4\rangle$ | $|-3\rangle$ | $|-2\rangle$ | $|-1\rangle$ | $|0\rangle$ | $|1\rangle$ | $|2\rangle$ | $|3\rangle$ | $|4\rangle$ |
|---|---|---|---|---|---|---|---|---|---|
| 0.000 | 0.0032 | 0.233 | -0.1466 | 0.6513 | 0 | 0.6513 | 0.1466 | 0.233 | -0.0032 |
| 0.215 | 0.048 | -0.0847 | 0.454 | -0.1168 | 0.7359 | 0.1168 | 0.454 | 0.0847 | 0.048 |
| 66.350 | 0.0804 | 0.5044 | 0.0175 | 0.4607 | 0.2303 | -0.4607 | 0.0175 | -0.5044 | 0.0804 |
| 71.962 | -0.1563 | -0.0045 | -0.6733 | -0.1491 | 0 | -0.1491 | 0.6733 | -0.0045 | 0.1563 |
| 125.759 | 0.1734 | 0.6456 | 0.0064 | -0.2304 | 0 | -0.2304 | -0.0064 | 0.6456 | -0.1734 |
| 139.848 | -0.2631 | -0.0842 | -0.4849 | 0.0029 | 0.6142 | -0.0029 | -0.4849 | 0.0842 | -0.2631 |
| 165.382 | -0.1318 | -0.4443 | 0.1136 | 0.5201 | -0.0602 | -0.5201 | 0.1136 | 0.4443 | -0.1318 |
| 223.947 | 0.6675 | -0.1699 | -0.1586 | 0.0218 | 0 | 0.0218 | 0.1586 | -0.1699 | -0.6675 |
| 228.163 | -0.6361 | 0.1842 | 0.2134 | -0.0595 | -0.1568 | 0.0595 | 0.2134 | -0.1842 | -0.6361 |

Interestingly, we observe a very small splitting, $\Delta E \approx 0.215$ meV, between the lowest two CEF levels $|\psi_+\rangle$ and $|\psi_-\rangle$. These two states are well separated from all other CEF levels, which are at least 60 meV in energy above. Therefore, $|\psi_\pm\rangle$ $ form a quasi-degenerate non-Kramers doublet. Certainly, our point-charge model calculations can be further refined via CEF parameter fitting, once either inelastic neutron scattering or Raman scattering results are available. But we expect the qualitative picture, which is already able to provide a reasonable understanding to experimental findings in $Pr_2Ga_2BeO_7$, will not change.

We can then define effective $S = 1/2$ spin operators associated with the quasi-doublet:

$$S^x = (|\psi_+\rangle\langle\psi_+| - |\psi_-\rangle\langle\psi_-|)/2, \tag{1}$$

$$S^x = -(|\psi_+\rangle\langle\psi_{-+}| - |\psi_-\rangle\langle\psi_{+-}|)/2i, \tag{2}$$

$$S^x = (|\psi_+\rangle\langle\psi_-| + |\psi_-\rangle\langle\psi_+|)/2. \tag{3}$$

One can check that these operators satisfy the $SU(2)$ Lie algebra. However, given the non-Kramers nature of the quasi-doublet, they transform differently under time reversal symmetry (TRS). By projecting the total angular momentum operators **J** into this quasi-doublet subspace, we find $S^z$ transforms as a magnetic dipole, while $S^x$ and $S^y$ transform as quadrupolar operators that preserve the TRS.

Considering the strong SOC, the exchange interactions are highly anisotropic in both real and spin spaces. Generically, the exchange coupling $\overleftrightarrow{J}$ is a bond-dependent tensor containing Kitaev-like and off-diagonal terms. This makes the effective model for $Pr_2Ga_2BeO_7$ very complicated, and we defer a full description and analysis on such a model in a later theoretical study[S16].

Here we consider a much simpler minimal model, which we show already captures the essential features of magnetic properties of $Pr_2Ga_2BeO_7$ in low temperatures. This is an XXZ-type spin model defined on a 2D SSL, and the Hamiltonian read as

$$H = \sum_{i,\delta} J_\delta [\Delta(S_i^x S_{i+\delta}^x + S_i^y S_{i+\delta}^y) + S_i^z S_{i+\delta}^z] - \sum_i (h_x S_i^x + h_v^z S_i^z), \qquad (4)$$

where $S_i^\alpha$ ($\alpha = x, y, z$) refers to the $S = 1/2$ spin operator associated with the doublet, $J_\delta = J_1, J_2$ denote the nearest and next nearest neighboring exchange couplings on the SSL, $\Delta$ is the spin anisotropic factor. $h_v^z$ is the applied magnetic field along the crystalline $v = a, c$ direction, and $h_x$ is the CEF splitting between $|\psi_+\rangle$ and $|\psi_-\rangle$. As mentioned above, only $S_i^z$ transforms as a magnetic dipole. Therefore, the applied magnetic field couples to $S_i^z$ only despite its direction. The average $J_\delta$ can then be estimated from the CW behavior of the susceptibility data. The extracted CW temperature $\theta_{CW} \sim -20$ K and the observation of no magnetic ordering down to 0.1 K indicate that the system is strongly frustrated. Moreover, this implies that the CEF splitting $h_x$ should not dominate over the exchange couplings.

Because $S^x$ and $S^y$ are non-magnetic quadrupoles, they cannot couple to the magnetic field linearly. It is then hard to determine the anisotropic factor $\Delta$ from existing magnetization and susceptibility data. Nonetheless, we expect that the model has a substantially strong XY-type spin anisotropy, e.g. $\Delta > 1$, for the following reasons. First, as shown in Table SIV, $|\psi_\pm\rangle$ are dominated by $|J^z = 0\rangle$ and $|J^z = \pm 1\rangle$ states, respectively. Quadrupolar interactions associated with processes that modify $\mathcal{J}^z$ by $\pm 1$ can be sizable. Note that this situation is completely different from some strong Ising anisotropic systems. There, the doublet is dominant by states with $\mathcal{J}^z = \pm J$ values where $J \gg 1$[S17-S20]. The quadrupolar interaction must come from high-order perturbation modifying $\mathcal{J}^z$ by a large number and is hence vanishingly small. As a second reason, a system with strong Ising anisotropy should exhibit 1/3 magnetization plateau under magnetic field[S21], which is not observed experimentally in $Pr_2Ga_2BeO_7$.

We then study the ground states of the XXZ Shastry-Sutherland model (SSM) in Eq. (4). In the Heisenberg SSM ($\Delta = 1$), it is well known that the ground state evolves from dimer singlets to a plaquette valence bond solid (PVBS) then to an AFM state by increasing $J_1/J_2$. The PVBS appears as the strong competing effect between $J_1$ and $J_2$. The quantum phase transition between the PVBS and the AFM state has been extensively studied by using various numerical techniques. It is agreed that the transition is very close to a deconfined quantum critical point (DQCP). However, it is still controversial whether an intervening quantum spin liquid (QSL) exists.

Here we study the evolution of the ground state of the XXZ SSM with varying the spin anisotropy $\Delta$. With quantum fluctuations tuned by the spin anisotropy, it is hoped that a QSL can be stabilized. Given the strong frustration of the material, we take $J_1/J_2 = 0.685$, where the Heisenberg model has a PVBS ground state. We calculate the ground state of the XXZ SSM by using the density matrix renormalization group (DMRG) method[S22]. To check whether the system is ordered. We have calculated the order parameters of the PVBS and AFM states, $m_p$ and $m_s$, respectively:

$$m_p = \frac{1}{N} \left| \sum_{ij \in \square_A} \langle \mathbf{S}_i \cdot \mathbf{S}_j \rangle - \sum_{ij \in \square_B} \langle \mathbf{S}_i \cdot \mathbf{S}_j \rangle \right|, \qquad (5)$$

$$m_s = \frac{1}{N} \sqrt{\sum_{\alpha \in x,y,z} \langle \sum_{ij} e^{i\pi r_i} S_{r_i}^\alpha \rangle^2}. \qquad (6)$$

As shown in Fig. S8(a), by increasing $\Delta$, the PVBS order parameter decreases and is vanishing for $\Delta \gtrsim 2.5$. Interestingly, at least for $\Delta \lesssim 3$ the AFM order parameter is also vanishing. The absence of both PVBS and AFM orders suggests a QSL is stabilized in this parameter regime. We further calculated the real-space spin correlation function $\langle \mathbf{S}_0 \mathbf{S}_r \rangle$ of this QSL-like state. As shown in Fig. S8(b), for $\Delta = 2.5$, the spin correlation function decays with spatial distance $r$ in an algebraic way. This indicates that the QSL-like state we found is gapless. As discussed in the main text, a gapless QSL naturally explains a serial of experimental findings in $Pr_2Ga_2BeO_7$, including the absence of magnetic order and the power-law temperature dependence of specific heat and thermal conductivity.

Theoretically, the nature of this QSL state needs to be clarified. Experimentally, the $T^2$ dependence of the specific heat indicates the elementary excitation has a linear dispersion. One likely scenario is that the ground state is a Dirac QSL with a linearly dispersive spinon excitations. In this case, disorder in the system may act as a random potential for spinons, which gives rise to a small but finite residual density of states at low energy. This well explains the finite $\kappa_0/T$ term in the thermal conductivity. Moreover, we fit the specific data between 0.2 K and 4 K with $\gamma T + aT^2$. As shown in the insets of Fig. S3, the finite $\gamma$ verifies the existence of a residual density of states.

As an alternative scenario, the ground state may also be a *d*-wave pairing $Z_2$ QSL. The spinon excitations around the nodes has linear dispersion in a way similar to a Dirac QSL. In this case, disorder can also induce a finite residual density of states and cause a universal $\kappa_0/T$ term, as well understood in high-$T_\text{c}$ cuprates[S23].

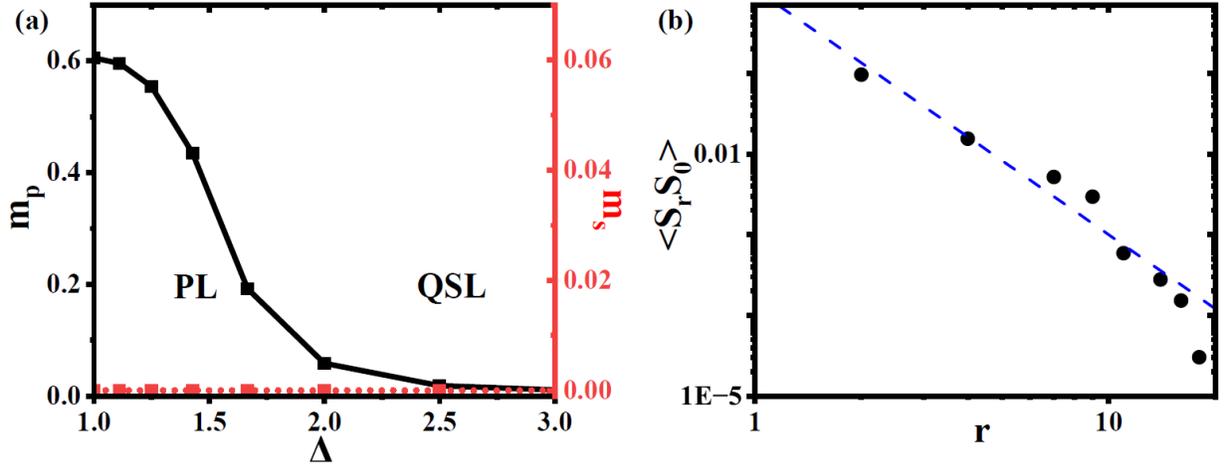

**Supplementary Figure S8** (a) Evolution of PVBS and AFM order parameters, $m_p$ and $m_s$, respectively, with $\Delta$ in the XXZ SSM defined in Eq. (4) at $J_1/J_2 = 0.685$ from a DMRG calculation. A QSL-like state with neither order is stabilized for $2.5 \lesssim \Delta \lesssim 3$. Note that for $\Delta > 1$, the AFM order in the effective model corresponds to an antiferro-quadrupolar order. (b) Real-space spin correlation of the QSL-like state at $\Delta = 2.5$ and $J_1/J_2 = 0.685$. The dashed line is a power-law fit. The algebraic decay of the spin correlation function with the distance $r$ suggests that the state is gapless.

## References

S1. Yamashita, M. *et al.* Thermal-transport measurements in a quantum spin-liquid state of the frustrated triangular magnet $\kappa$-(BEDT-TTF)$_2$Cu$_2$(CN)$_3$, *Nat. Phys.* **5**, 44-47 (2009).

S2. Yamashita, S. *et al.* Thermodynamic properties of a spin-1/2 spin-liquid state in a $\kappa$-type organic salt. *Nat. Phys.* **4**, 459-462 (2008).

S3. Han, T.-H. *et al.* Fractionalized excitations in the spin-liquid state of a kagome-lattice antiferromagnet. *Nature* **492**, 406-410 (2012).

S4. Helton, J. S. *et al.* Spin Dynamics of the Spin-1/2 Kagome Lattice Antiferromagnet ZnCu$_3$(OH)$_6$Cl$_2$. *Phys. Rev. Lett.* **98**, 107204 (2007).

S5. Huang, Y. Y. *et al.* Heat Transport in Herbertsmithite: Can a Quantum Spin Liquid Survive Disorder? *Phys. Rev. Lett.* **127**, 267202 (2021).

S6. Zeng, Z. *et al.* Possible Dirac quantum spin liquid in a kagome quantum antiferromagnet YCu$_3$(OH)$_6$Br$_2$[Br$_x$(OH)$_{1-x}$]. *Phys. Rev. B* **105**, L121109 (2022).

S7. Hong, X. *et al.* Heat transport of the kagome Heisenberg quantum spin liquid candidate YCu$_3$(OH)$_{6.5}$Br$_{2.5}$: Localized magnetic excitations and a putative spin gap. *Phys. Rev. B* **106**, L220406 (2022).

S8. Jeon, S. *et al.* One-ninth magnetization plateau stabilized by spin entanglement in a kagome antiferromagnet. *Nat. Phys.* **20**, 435-441 (2024).

S9. Dai, P.-L. *et al.* Spinon Fermi Surface Spin Liquid in a Triangular Lattice Antiferromagnet NaYbSe$_2$. *Phys. Rev. X* **11**, 021044 (2021).

S10. Zhang, Z. *et al.* Low-energy spin dynamics of the quantum spin liquid candidate NaYbSe$_2$. *Phys. Rev. B* **106**, 085115 (2022).

S11. Zhu, Z. *et al.* Fluctuating magnetic droplets immersed in a sea of quantum spin liquid. *The Innovation* **4**, 100459 (2023).

S12. Balz, C. *et al.* Physical realization of a quantum spin liquid based on a complex frustration mechanism. *Nat. Phys*. **12**, 942-949 (2016).

S13. Sonnenschein, J. *et al.* Signatures for spinons in the quantum spin liquid candidate Ca$_{10}$Cr$_7$O$_{28}$. *Phys. Rev. B* **100**, 174428 (2019).

S14. Ni, J. M. *et al.* Ultralow-temperature heat transport in the quantum spin liquid candidate Ca$_{10}$Cr$_7$O$_{28}$ with a bilayer kagome lattice. *Phys. Rev. B* **97**, 104413 (2018).

S15. Scheie, A., PyCrystalField: Software for Calculation, Analysis, and Fitting of Crystal Electric Field Hamiltonians. *J. Appl. Cryst.* **54**, 356-362 (2021).

S16. Duan, G., Liu, C., Yu, R., in preparation.

S17. Siemensmeyer, K. *et al.* Fractional magnetization plateaus and magnetic order in the Shastry-Sutherland magnet TmB$_4$. *Phys. Rev. Lett.* **101**,177201 (2008).

S18. Nagl, J. *et al.* Excitation Spectrum and Spin Hamiltonian of the Frustrated Quantum Ising Magnet Pr$_3$BWO$_9$. arXiv:2402.14107 (2024).

S19. Shen, Y. *et al.* Intertwined dipolar and multipolar order in the triangular-lattice magnet TmMgGaO$_4$. *Nat. Commun.* **10**, 4530 (2019).

S20. Dun, Z. *et al.* Effective point-charge analysis of crystal fields: Application to rare-earth pyrochlores and tripod kagome magnets *R*$_3$Mg$_2$Sb$_3$O$_{14}$. *Phys. Rev. Res*earch **3**, 023012 (2021).

S21. Dublenych, Y. I. Ground states of the Ising model on the Shastry-Sutherland lattice and the origin of the fractional magnetization plateaus in rare-earth-metal tetraborides. *Phys. Rev. Lett.* **109**, 167202 (2012).

S22. White, S. R. Density matrix formulation for quantum renormalization groups. *Phys. Rev. Lett.* **69**, 2863-2866 (1992).

S23. Hussey, N. E. Low-energy quasiparticles in high-$T_c$ cuprates. *Adv. Phys.* **51**, 1685-1771 (2002).